\documentclass[11pt]{article}
\usepackage{a4wide}
\usepackage{amssymb}
\usepackage{amsmath}
\usepackage{amsthm}
\usepackage{natbib}

\newtheorem{theorem}{Theorem}

\newtheorem{assumption}[theorem]{Assumption}

\newtheorem{definition}[theorem]{Definition}

\newtheorem{lemma}[theorem]{Lemma}

\newtheorem{proposition}[theorem]{Proposition}

\newtheorem{remark}[theorem]{Remark}

\newtheorem{discussion}[theorem]{Discussion}

\newcommand{\diag}{\mathop{\mathrm{diag}}}

\newcommand{\dom}{\mathop{\mathrm{dom}}}
\newcommand{\Var}{\mathop{\mathrm{Var}}}

\def\var{\text{Var}}

\begin{document}

\title{Statistical Arbitrage and Optimal Trading with Transaction Costs in Futures Markets}

\author{Theodoros Tsagaris\footnote{Imperial College London, Department of Mathematics, and BlueCrest Capital Management. The views presented here reflect solely the author's opinion.
 }}
\maketitle

\begin{abstract}
We consider the Brownian market model and the problem of expected utility maximization of terminal wealth. 
We, specifically, examine the problem of maximizing the utility of terminal wealth under the presence of transaction costs of a fund/agent investing in futures markets.  We offer some preliminary remarks about statistical arbitrage strategies
and we set the framework for futures markets,
and introduce concepts such as margin, gearing and slippage. The setting is of discrete time, and the price evolution of the futures prices is modelled as discrete random sequence involving Ito's sums. We assume the drift and
the Brownian motion driving the return process are non-observable and the
transaction costs are represented by the bid-ask spread. We provide explicit solution to the optimal portfolio process, and we offer an example using
logarithmic utility.
\end{abstract}


\section{Introduction}

\subsection{Optimal Investment}
This section examines the problem of a fund/agent with assets under management
$x$, who wishes to invest in $d$ assets, and aims to maximize the expectation
of the utility function of the terminal wealth $X_{t_N}$.
The agent aims to invest her initial capital on $d$-dimensional futures contracts,
which are driven by a $d$-dimensional Brownian motion. The manager/investor has only
partial information, in the sense that she neither observes the drift $\beta$,
not the Brownian motion $W$. In addition, the investor incurs a transaction
cost equal to half the bid/ask spread $c$ for each lot(contract) traded.
The investor's objective is to achieve an optimal position ($P_{t_n}$) process
that can hedge against the risk that is associated with the market, and deliver
a constant level variance of the portfolio returns taking into consideration
in the wealth equation any capital that is not required by margin calls.
Finally, the manager would like to have the option to invest in a risk neutral
manner i.e. the correlation between the underlying asset prices and the investment
returns is 0.

The following text is structured as follows:

We proceed in \ref{subsec:StatArb} with some explanatory remarks on relative/statistical arbitrage trading strategies commonly employed by proprietary desks. We then
continue \ref{subsec:MarketModel} on the definition of the futures price evolution process and the conditions under a unique and sufficient solution exists. We operate in the usual complete probability space  $(\Omega,\mathcal{F},(\mathcal{F})_{t_{n\geq0}},P)$, and our processes are adapted to the augmented filtration $\mathcal{F}_{t_n}$ generated by non-decreasing $\sigma$-algebras. We define in addition
the filtration generated by the futures prices and geometric returns $\mathcal{F}^F_{t_n}$,
$\mathcal{F}^R_{t_n}$, respectively and show by direct observation that are
equal.

In section \ref{subsec:TradingStrat}, we offer a remark of how to construct
a statistical arbitrage strategy using the innovations approach, a tool very commonly used in stochastic filtering theory (see \cite{O06}). Then, we continue with \ref{subSec:InvProcess}, where we offer necessary definitions and remarks on the operation of futures funds. We introduce some essential notions for the derivation of the optimal trading strategy such as gearing, position, unit value, contract size, margin and slippage.
In \ref{subSec:ProblemFormulation}, we define the relative risk process under
transaction costs, and determine a new measure under which the expected value
of the discounted final wealth is equal to the initial assets under management.
We introduce a new formulation for the wealth process, and show explicitly
from first principles how the wealth stochastic difference equation can be constructed using actual
(bid/ask spread) rather than proportional transaction
costs to the wealth process. We obtain an explicit solution for the cost term (see eq. (\ref{eq:costFn}))
$$\widetilde{c}^{(i)}_{t_n}=\frac{c^{(i)}f^{(i)}|\Delta P^{\pi,c(i)}_{t_{n-1}}|}
{2P^{\pi,c(i)}_{t_n}C^{(i)}_{t_n}\Delta t}$$
and gain the insight that the cost term
tends to infinity as the partition $\Pi=\max_t \Delta t_n$. 
tends to 0.
We also
introduce the adjusted discount process as it has been amended for margin
calls.

For the remainder of the paper, we make use of the  auxiliary results, presented
in appendix \ref{sec:auxResults}, regarding
concave functional evaluated in some random process realization, and we state
some known results about utility functions.
We continue with section \ref{sec:OptimProblem} where we establish the primal optimization problem
and its dual.
The main result is presented in section \ref{sec:OptimPortf}, where we prove
the optimality of the portfolio process
$\pi$ in a similar manner to \cite{KLS87}, and we characterize the optimal portfolio process using the martingale representation theorem.
Finally, in section \ref{sec:Examples}, we offer an example using logarithmic
utility, and show explicitly that the terminal wealth maximizes the value
function. Without sacrificing generality, and in order to offer an explicit
solution to the reader, we suggest an approximating optimal portfolio process by using
an approximate functional $\widehat{c}$ for the cost term $\widetilde{c}$.
A remark is finally made that the investor should be interested in solution
that satisfy the feasibility condition $|\widehat{\beta}|-|\widetilde{c}|\geq0$.

\subsection{Previous Work}
The theory of optimal investment dates back to the seminal paper of Markowitz\index{Markowitz}
\cite{M52} whose idea based on the examination of effective diversification
of an investor's monetary resources. 
In the seminal papers of Merton \cite{M70,M69}, a solution to the problem of optimal investment is presented for complete markets in the continuous
case using tools of optimal stochastic control. Merton uses
Hamilton-Jacobi-Bellman equation to derive a partial differential equation
for the special case where asset prices are Markovian. 

In 1973, Bismut \cite{B75} examines the optimal investment problem using
a conjugate function in an optimal stochastic control setting. 
Bismut contribution 
forms what is known now as the martingale approach in convex duality theory. Harrison and Pliska \cite{HP81} provide
a generic market model in continuous market trading and show that the optimal
control problem can be solved using the martingale representation characterization
of the terminal wealth by investing in complete markets. Cox, Huang \cite{CH88}
present sufficient conditions for the existence of solution to the
consumption-portfolio problem under a certain class of utility functions. Karatzas, Lehoczky and Shreve \cite{KLS87} split the optimal investment problem into
two more elementary problems where a small investor (one who cannot affect the market through his transactions) attempts to maximize the utility
of consumption and terminal wealth separately.

Further studies by Cvitanic, Karatzas \cite{CK92} generalize the results
to the constrained case where the portfolio takes values in a predetermined
closed, convex subset. Cuoco \cite{C97} extends the latter problem to the
case of stochastic initial endowment. Karatzas, Lehoczky, Shreve and Xu
\cite{KLSX91} study the case where the number of stocks is strictly smaller
than the dimension of Brownian motions (incomplete markets) and suggest a solution by introducing additional fictitious stocks to complete the market. Xu, Shreve \cite{XS92} consider the finite horizon problem of optimal consumption/portfolio
under short selling prohibition.

Kramkov, Schachermayer \cite{KS99}  study the problem of optimal investment
in the  more general framework of semimartingales, and provide necessary
and sufficient conditions for the solution to the maximization of the expected
utility of terminal wealth. They introduce the asymptotic elasticity of the
utility function and show that it is necessary and sufficient condition for the optimal solution if it is strictly less than $1$.

Transaction costs are an important financial feature whose omission overestimates
the wealth process, and subsequently affects the theoretical or derivative
prices of the underlying assets. Davis, Norman \cite{DN90} have showed that the optimal policies are local times at the boundary, and that the boundary is determined by a nonlinear free boundary problem. Shreve, Soner \cite{SS94} are influenced by Davis, Norman
work to provide an analysis based on viscosity solutions to Hamilton-Jacobi-Bellman
equation and additionally show that for a power utility $c^p/p, \ 0<p<1$
there is a correspondence between the value function and transaction cost
in the order of $2/3$ power. Zariphopoulou \cite{Z92} shows that the value
function is a viscosity solution, in the sense of Crandall, Lions \cite{CL83},
of a system of variational inequalities with gradient constraints. Cvitanic,
Karatzas \cite{CK96} provide an analysis of hedging a claim under transaction
costs  using duality theory. Cvitanic, Pham, Touzi \cite{CPT99} apply \cite{CK96}
to show that the cheapest strategy to hedging is a buy and hold strategy
of a call option. Kabanov, Last \cite{KL02} deploy the more general semimartingale
model with transaction costs in a continuous setting. Under assumption of
properness of the solvency cone, they provide characterization of the initial
endowments that allow investors to hedge contingent claims.

Lakner \cite{L98} approaches the optimal investment problem from the partial
information case. He deploys some classical theory (Kalman-Bucy) to find
an explicit solution for the portfolio process when the drift process is
the unobservable component. Pham, Quenez \cite{PQ01} use stochastic duality
and filtering theory to characterize the optimal portfolio process subject
to the assumption the stock prices follow a stochastic volatility model.

\subsection{Market Neutral Statistical Arbitrage Trading Strategies}\label{subsec:StatArb}
Statistical arbitrage methodologies refer to investment strategies
that have high probability of outperforming the market (see
notes in chapter 4 of \cite{S04}). A very common, and particularly attractive to investors,
statistical arbitrage strategy is called market-neutral strategy. Funds that
deploy market-neutral strategies 
(sometimes refer to stat. arb. funds) typically pledge to investors to
deliver portfolio returns that are uncorrelated on average
with the underlying asset or sector in question.

Market neutral strategies are generally multivariate techniques (or
bivariate in the special case of pairs trading) to define estimated
equilibrium levels for a basket of assets, one may arbitrarily
deploy. Such techniques  generate a synthetic asset based on certain
variables that could be interpreted as the fair price for a subset of
factors from the available universe. Commonly used variables
are other assets classified in the same sector/industry or macroeconomic factors.
Deviations from the fair price are considered as mis-pricings from the
equilibrium price and one could use an arbitrary trading strategy to
exploit such seemingly opportunity patterns. The forecasting ability
of these strategies lies in the
mean-reverting dynamics of the innovation process (for statistical arbitrage
related remarks/techniques and an application, see \cite{Montana2007}).

In this study, we motivate our
methodology on market neutral strategies and we offer some explanatory remarks
of how such strategies can be constructed using standard filtering techniques.
These techniques provide estimates of drift that are uncorrelated to the
underlying price, which in turn are deployed in the optimal portfolio characterizations
derived by classical duality theory in optimal investment.

\section{The model}
\subsection{The Market Model}\label{subsec:MarketModel}
Assume we are given a complete probability space
$(\Omega,\mathcal{F},(\mathcal{F})_{t_n\geq0},P)$ , and let $\mathcal{F}_{t_n}$ be the augmented
filtration generated by the union of a non-decreasing family of
sub-$\sigma$-algebras and $\mathcal{N}$-null sets.

Let $\mathbf{T}$ be a finite closed interval $[0,T]$, where $T$ is the terminal time.
Assume we have the following partition
$0=t_0<t_1<\ldots<t_n\ldots<t_N=T$
of the set $\mathbf{T}$, and we denote the set of non-decreasing elements
by $\mathcal{S}=\{t_0,t_1,\ldots,t_n,\ldots,t_N\}$. 
Moreover, we introduce the following notation for the time increment
$\Delta t=\Delta t_n=t_{n+1}-t_n$, and assume equidistant increments 
$t_n=t_0+n\Pi$, where $\Pi$ is the maximum timestep 
$\Pi=\max_t \Delta t_n$.

We introduce the process $F_{t_n}$ as a mapping from
$\Omega \times \mathcal{S}\rightarrow \mathbb{R}^d$, for which is measurable with
respect to the smallest-sigma field of subsets of $\Omega \times \mathcal{S}$.
In this setting, there is no meaning to assume right continuity.
The \emph{optional} $\sigma$-field $\mathcal{O}$ is the $\sigma$-field generated
by all adapted processes to $(\mathcal{F}_{t_n})_{n \in \mathbb{N}}$, such that
$F_{t_n}$ is $\mathcal{F}_{t_n}$ measurable for each $n\in \mathbb{N}$.

Denote by
\begin{equation}\label{eqn:filtration}
\mathcal{F}^F_{t_n} = \sigma \left( F_{t_n}; t_n \in \mathcal{S}, n \in \mathbb{N}\right) 
\end{equation} 
the  $\sigma$-field generated by the futures contracts prices.
The sigma field $\mathcal{F}^F_{t_n}$ is a subfield of 
$\{\mathcal{F}_{t_n}; t_n \in \mathcal{S}, n \in \mathbb{N}\}$. The filtration $\mathcal{F}_t^F$ represents the information that is available
to investors up to, and including time $t_n$.

We adopt a financial market consisting of $d$-futures $F^{(i)}(\omega,t_n), \
i=1,...,d, \ \forall t_n \in \mathcal{S}$ that 
satisfies the following evolution
equation\footnote{Henceforth, we also make use of the following
condensed notation: $f_{t_n}=f(\omega,t_n)$.}
\begin{equation}\label{futEvol}
\Delta F^{(i)}_{t_n}=F^{(i)}_{t_n}\left(\beta^{(i)}( t_n)\Delta t+\sum_{j=1}^d\sigma^{(ij)}
\Delta W^{(j)}_{t_n}\right)
\end{equation}
where
\begin{itemize}
    \item $W_{t_n}$ is a $d$ dimensional Brownian motion with correlation
    matrix $\rho$ i.e. 
    $E[\Delta W_{t_n}\Delta W^*_{t_n}]=\rho \Delta t$.

    $\rho= \rho^{(ij)},\ i=1,\dots,d ,j=1,\dots,d $ is
    a $d\times d$ non-random  matrix satisfying $
    \rho^{(\cdot,\cdot)} < \infty$ (where $(\cdot)$ denotes
    summation over all subscripts)
    \item $\beta_{t_n}$ is a $d$ dimensional mean rate of return process
    subject to $\sum_{k=0}^{n-1}||\beta_{t_k}||\Delta t< \infty$.
     $\beta_{t_n}:\Omega \times \mathcal{S} \rightarrow R^d$
is a measurable mapping such that  $\beta_{t_n}$ is 
$\mathcal{F}_{t_n}$ measurable for every $t_n\in \mathcal{S}$.
    \item $\sigma= \sigma^{(ij)},\ i=1,\dots,d ,j=1,\dots,d $ is
    a $d\times d$ non-random volatility matrix satisfying 
    $\sigma^{(\cdot,\cdot)} < \infty$ 
    \item The initial prices $F^0$ is a d-dimensional vector of deterministic
    constants
\end{itemize}
and the following statements hold
$$\mathcal{F}^{F^{(0)},F}_{t_n} \subseteq \mathcal{F}_{t_n}$$   
and
$\mathcal{F}_{t_n}$ is independent of $\sigma(\Delta W_{t_n})$.

\begin{assumption}\label{assumption:nonsingular}
We assume $\sigma$ and $\rho$ have full rank, therefore are nonsingular and hence invertible. (See Lemma 2.1 of Karatzas, Lehoczky, Shreve (1987) on
uniform ellipticity as sufficient condition for invertibility )
\end{assumption}

By Karatzas and Shreve, eq.  (\ref{futEvol})
has a weak solution (see \cite{KS91}, Chapter 5, Problem 6.15).

All vectors are column vectors, unless stated otherwise. We will
denote the transpose by $*$ (i.e. $(v^*)^*=v)$.

We assume that the agent can only observe the futures prices but the drift process and the Brownian motion $W$ are unobservable. The initial prices
$F^{(0)}$ and the volatility matrix $\sigma$ are known in advance.

Moreover, we denote the quadratic variation of $W_{t_n}$ by 
$$[W,W]_n=\sum_{k=0}^{n-1} (W_{t_{k+1}}-W_{t_k})(W_{t_{k+1}}-W_{t_k})^*$$
\begin{assumption}
We will be making use, throughout the remainder of the  text, of the simplifying assumption that the the quadratic variation can be replaced by its
expectation.
\end{assumption}

Finally, throughout the text, we
assume that any sequence of measurable mappings $Y_n, \ n \in
\mathbb{N},$ takes values in a Polish space $(E,\mathcal{B})$ and
hence we have that for any Borel function $h:E \rightarrow
\mathbb{R}$ , the function $h(Y_n)$ converges to $h(X)$ surely. The
above assumption is crucial for the existence of conditional
probability of a mapping of $Y$ given a sub-$\sigma$-algebra
$\Sigma\subseteq\mathcal{B}$ (see Stroock,Varadhan 2000).

\subsection{The Trading Strategy}\label{subsec:TradingStrat}

Define the $d$-dimensional return process $R:\mathcal{S}
\rightarrow\mathbb{R}_+^{d}$ by
\begin{equation}\label{retEvol1}
\Delta F^{(i)}_{t_n}=F^{(i)}_{t_n}\Delta R^{(i)}_{t_n}
\end{equation}
and applying (\ref{futEvol}) in the above, we obtain

\begin{equation}\label{retEvol2}
\Delta R^{(i)}_{t_n}=\beta^{(i)}_{t_n}\Delta t+\sum_{j=1}^d\sigma^{(ij)} \Delta W^{(j)}_{t_n}
\end{equation}

\begin{lemma}
The filtration generated
by the return process
$$\mathcal{F}^{R}_{t_n}=\sigma\{R_{t_1},\ldots,R_{t_n}\} \bigvee \{P-null \  sets \ of  \ \mathcal{F}\} $$
is equal to the augmented filtration defined in (\ref{eqn:filtration})
$\mathcal{F}^F_{t_n}=\mathcal{F}^R_{t_n}$
\end{lemma}

\begin{proof}
Eq. (\ref{retEvol1}) implies that both
$\mathcal{F}^{R}_{t_n}$, $\mathcal{F}^{F}_{t_n}$ generate the same filtration.
\end{proof}

From the above lemma and for the remainder of the text, we take  the return process $R=(R_{t_n},\mathcal{F}^R_{t_n})$ as the observable random process, and $\beta=(\beta_{t_n},\mathcal{F}_{t_n})$ as the unobservable component.
\begin{assumption}
The linear dynamics of $\beta_{t_n}$ are given by the d-dimensional
process
\begin{equation}\label{eqn:retDriftEvol}
\Delta\beta^{(i)}_{t_n}=\sum_{j=1}^d\alpha^{(ij)}\beta^{(j)}_{t_n}\Delta t+\sum_{j=1}^d\varsigma^{(ij)}\Delta W^{(2)j}_{t_n}
\end{equation}
where $W^{(2)}$ is a $d$-dimensional Brownian motion and  $W$, $W^{(2)}$ are independent. The matrices $\alpha$ and $\varsigma$ are non-random
in $R^{d \times d}$ and satisfy the following conditions

$$\sum_{k=0}^{n-1} E[\|\alpha\|\Delta t] < \infty, \ \ \sum_{k=0}^{n-1} E[\|\varsigma\|\Delta t] < \infty$$

\end{assumption}

$\beta_{t_0}$ is the initial condition satisfying 
$ E[\|\beta_{t_0}\|\Delta t] < \infty$ and is independent of 
$\sigma\{W_{t_1},\ldots,W_{t_n}\}$ and $\sigma\{W^{(2)}_{t_1},\ldots,W^{(2)}_{t_n}\}$.

Moreover, 
$\mathcal{F}_{t_n}$ is independent of $\sigma(\Delta W^{(2)}_{t_n})$.

This is a classical partial information estimation problem of finding the
state process $\beta_{t_n}$ by observing the returns of the futures prices.
The investor only observes the futures contract prices, not the Brownian motion or
the drift. For a broad class of mean square cost functions, it can be
shown that the optimal estimate, in terms of minimising the expected value
of the cost functions, is the expectation conditional on the filtration generated
by the observation process.

Let $P_{\mathcal{N}}^{(i)}$ denote the projection from the Hilbert space $ L^2(P)$
onto the closure in $L^2(P)$ of linearly combined functions
$$\mathcal{N}=\{a_0+a_1 R^1_{t_n}+...+a_{d} R^{d}_{t_n} \in L^2(P) \  ; \  R \  is \  \mathcal{F}^R_{t_n}
\  measurable \}$$
Then, $P_{\mathcal{N}}^{(i)}$ coincides with the conditional expectation

\begin{equation}\label{eqn:projection}
P_{\mathcal{N}}^{(i)}=E[\beta^{(i)}_{t_n} | \mathcal{F}^R_{t_n}]=\widehat{\beta}^{(i)}_{t_n}
\end{equation}

(see \cite{B92,D84,Kall80,K84,K90,LS00a,O06}
for proof of this known result). We denote by $\beta_{t_n}$
the vector with elements 
$\{ \beta^1_{t_n},\beta^2_{t_n},\ldots,\beta^d_{t_n}\}$.

Then, the innovation process is defined as

\begin{equation}\label{eqn:innovProcess1}
\Delta\nu_{t_n}=(\sigma)^{-1}(\Delta R_{t_n}-\widehat{\beta}_{t_n}\Delta t)
\end{equation}

\begin{equation}\label{eqn:innovProcess2}
\Delta\nu_{t_n}=(\sigma)^{-1}(\beta_{t_n}-\widehat{\beta}_{t_n})\Delta t +  \Delta W_{t_n}
\end{equation}

\begin{lemma}\label{lemma:stratGains}
The conditional expectation $\Delta\nu_{t_n}$ is given by
$$E[\Delta \nu^{(i)}_{t_n}|\mathcal{F}_{t_{n-1}}^R]=0$$
and the conditional variance is of the order of
$$\Var[\Delta\nu^{(i)}_{t_n}|\mathcal{F}_{t_{n-1}}^R]=
\rho_{ii}(t_{n+1}-t_n) =\Delta t$$
\end{lemma}
\begin{proof}
The first statement follows form iterated conditional expectation
$$E[\nu_{t_n}|F^R_{t_{n-1}}]=E[\sum_{k=0}^{n-1}(\sigma)^{-1}(\beta_{t_k}-\widehat{\beta}_{t_k})\Delta
t
+\sum_{k=0}^{n-1}\Delta W_{t_k}|F^R_{t_{n-1}}]$$
$$E[\nu_{t_n}|F^R_{t_{n-1}}]=E[\sum_{k=0}^{n-1}(\sigma)^{-1}E[(\beta_{t_k}-\widehat{\beta}_{t_k})|F^R_{t_k}]|F^R_{t_{n-1}}]
+E[\sum_{k=0}^{n-1} E[\Delta W_{t_k}|F^R_{t_k}]|F^R_{t_{n-1}}]=0$$
By discarding the quadratic $\Delta t$ terms, and replacing the quadratic
variation by its expectation, we get
$$\Var[\nu_{t_n}|\mathcal{F}_{t_{n-1}}^R]=E[\nu^*_{t_n}\nu_{t_n}|\mathcal{F}_{t_{n-1}}^R]=
E[E[\sum_{k=0}^{n-1}\Delta [W^*_{t_k}W_{t_k}]|F^R_{t_n}]|F^R_{t_{n-1}}]$$
$$\Var[\nu_{t_n}-\nu_{t_0}|\mathcal{F}_{t_{n-1}}^R]=\sum_{k=0}^{n-1} \rho \Delta t$$
\end{proof}

\begin{remark}
From lemma (\ref{lemma:stratGains}), we deduce the mean reversion properties
of the innovation process $\Delta\nu$ around a constant level. The stationary nature
of the spread (innovation) has been widely exploited by proprietary traders
to establish appropriate trading rules to benefit from the "mis-pricings"
of the futures contract price. The futures contract price is commonly interpreted
as expensive when the spread is positive and cheap when negative.

The use of spread, for the establishment of proprietary trading strategies
or hedging, is particularly attractive because  is uncorrelated to the futures
prices. By construction, the spread  is uncorrelated
to futures prices because of the orthogonality condition of the projection
theorem. Any proprietary strategy the depends solely on the spread
is typically a "market neutral" strategy because it is not influenced by the
fluctuations in the underlying asset (market).

In reality though, the proprietary choice of filter to calculate $\widehat{\beta}$ leads to estimation issues: It introduces estimation error, as generally
is notoriously difficult to capture the drift rate accurately, because of
the low signal to noise ratio of financial data. And, moreover, the necessary
and sufficient condition of orthogonality, may not always hold, and hence
the strategy might be occasionally affected by market fluctuations.
\end{remark}

\subsection{The Investment Process}\label{subSec:InvProcess}

In futures markets, an agent enters the market by paying usually a
locally denominated amount that reflects the price of the contract
at time $t$. Generally, the price of the contract is calculated by
$C_{t_n}=\diag(f)F_{t_n}$ ,where $f \in \mathbb{R}_+^{d}$ represents
the \emph{unit value} of the futures contract
\footnote{The locally denominated amount that corresponds to the one unit
movement in the future market price.}
and $C:\mathcal{S}
\rightarrow\mathbb{R}_+^{d}$ is a measurable process with respect to
$\mathcal{F}_{t_n}$. The agent is required to set aside an initial and
maintenance margin $m \in [0,1]$ (henceforth to be taken as
constant), generally $20\%$ of the current market, unlike stocks
where the margin could be roughly as much as $50\%(\textrm{sale proceeds plus } 50\%)$ for long (short) positions.

\begin{definition}\label{def:pos}
Consider the mappings $\pi:\mathcal{S} \times \Omega \rightarrow
\mathbb{R}^d$ and $P^{\pi}:\mathcal{S} \times \Omega \rightarrow \mathbb{R}^d$.
Then, the wealth process  corresponding to the investment strategy $\pi$
and initial endowment $x$ is denoted by $X^{\pi,x}_{t_n}$, and represents the
income accumulated up to, and including time $t$. The portfolio or
investment process $\pi$ represents the portfolio weight assigned to asset
$i$ (or correspondingly the  position $P^{\pi}_{t_n}$ of futures contracts, as we will later establish). The portfolio process is measurable with respect
to $\mathcal{F}^R_{t_n}$, i.e. the investor's portfolio allocation is based on information received only by the futures prices, not the drift or the Brownian
motion.
\end{definition}

 In the following remark, we offer some explanatory comments for the portfolio
 process and the orderbook that would facilitate the build-up of the
 formulation in the following text.

\begin{remark}\label{remark:orderbook}
 Consider at any point of time $t$, there is an orderbook of
futures positions $\mathcal{P}^{(ij)}_{t_n}, j \in 1,\dots J$
available, that consists of a stack of positions that reflects the
bid and ask prices, where $j$ refers to the ordering
of the stack. We assume that the
position, offered for purchase or bid for sale at the top of the
stack $\mathcal{P}^{(i0)}_{t_n}$, represents the narrowest spread
$c\geq 0$. An investor whose preference is to take position
$P^{\pi(i)} _{t_n}>\mathcal{P}^{(i0)}_{t_n}$ will need to move deeper in to the
stack, and pay the additional slippage incurred.\footnote{For real
time examples of the futures orderbook, check for instance Chicago
Board Of Trade web site.}
\end{remark}

\begin{assumption}\label{maxPos}
The position we take in the market is bounded from above
$P^{\pi(i)}_{t_n}\leq \mathcal{P}^{(i0)}_{t_n}$, in the sense that we can not
exceed the available position that is associated with the narrowest
spread.
\end{assumption}

\begin{assumption}\label{slippage}
 The assume that half the bid/ask spread $c/2$ represents the deviation in market unit terms
 between the intended and the actual "fill" price for any buy or sell trade.
 For simplicity, we assume that $c$ is a constant and does not fluctuate during the
period of the normal trading hours.

The cost an investor incurs in the market is only equal to the slippage
and there is no commission cost associated with any trade. This assumption
is not restrictive as commission tends to be only  a small contribution to the
overall cost.
\end{assumption}

\begin{discussion}\label{remark-futTrad}
 Contrary to the literature, where the optimal
investment problem is set up on $d+1$ assets,(e.g. $d$ stocks and a
bank account), it is reasonable for futures investment to consider that
we invest the entire capital in risky assets, and that the proportion of capital $1-m,\  \forall m \in [0,1]$ that is not required by margin calls is deposited at the
risk-free rate. Note that the same capital is invested concurrently
on risky assets and risk-free deposit account.
\end{discussion}

Consider an investor who wants to invest in $d$ assets, and his
initial endowment is $x>0$. The portfolio process is an
$\mathcal{F}^R_{t_n}$-measurable adapted
process that satisfies

$$E[\sum_{k=0}^{n-1} || \pi_{t_k}||^2 \Delta t] < \infty \ \ a.s. $$
Note, that we allow for short or long position, and that we could
relax any constraint on the position, subject to the fact that
assumption (\ref{maxPos}) holds throughout this text.


\subsection{The Problem Formulation}\label{subSec:ProblemFormulation}

In this section, we use throughout for consistency the same notation as in \cite{KLS87},\cite{KLSX91},\cite{KS98},\cite{CK92}.

The next lemma is rather crucial for section continuity, as we establish
the new measure under the underlying probability space. The new measure adjusts for the drift in (\ref{wealthProcess}), and under
this measure the wealth equation will turn into a local martingale.
The transformation is attained through a representative of the Radon-Nikodym derivative
\begin{equation}\label{eqn:RadonNikodym}
d\widetilde{P}_n=Z_{t_n}dP_n
\end{equation}
where $Z$ is given in eq. (\ref{expMartingale}), and assume it satisfies the Novikov condition (\ref{Novikov}). Consequently, we establish the absolute continuity of measures $\widetilde{P}_n<<P_n$, and we deduce that
$$\widetilde{W}_{t_n}=W_{t_n}+\sum_{k=0}^{n-1}\rho\theta_{t_k}\Delta t_k$$
is a Brownian motion under the new measure $\widetilde{P}_n$.

\begin{definition}\label{marketPriceOfRiskDef}
 Consider an investor who operates in the futures markets. The \emph{relative risk process under transaction costs} is a
 progressively measurable process $\theta_{t_n}:\mathcal{S}\rightarrow \mathbb{R}^d$
 that satisfies
\begin{equation}\label{marketPriceOfRisk}
\rho\sigma\theta_{t_n}=\beta_{t_n}-\widetilde{c}_{t_n}
\end{equation}
\end{definition}

\begin{definition}
Let $Z_n$ be a $P_n$-martingale given by
\begin{equation}\label{expMartingale}
Z_{t_n}=\exp\{-\sum_{k=0}^{n-1}
\theta^*_{t_k}\Delta W(s)-\frac{1}{2}\sum_{k=0}^{n-1} \theta^*_{t_k}\rho\theta_{t_k}\Delta
t  \}
\end{equation}
We denote the optional projection of $Z_n(t)$ to $\mathcal{F}^R_{t_n}$
by
$$\zeta_{t_n}=E[Z_{t_n}|\mathcal{F}^R_{t_n}]$$
\end{definition}

\begin{lemma}\label{GirsanovLemma}
Let
\begin{equation}\label{brownianUnderRiskNeutral}
\widetilde{W}_{t_n}=W_{t_n}+\sum_{k=0}^{n-1}\rho\theta_{t_k}\Delta t_k
\end{equation}

where $\theta^*_{t_n}\rho\theta_{t_n}$ satisfies the following condition

$$P[\omega: \sum_{k=0}^{n-1}\theta^*_{t_n}\rho\theta_{t_n} \Delta t_k<\infty]=1$$

If $E[Z_{t_n}]=1$, then $Z_{t_n}$ is a local martingale and a supermartingale and there is another probability
measure $\widetilde{P}_n$ that is $\mathcal{F}_{t_n}$-adapted, and corresponds to the
Radon-Nikodym derivative $d\widetilde{P}_n=Z_{t_n}dP_n$ and $P_n<< \widetilde{P}_n$.
The process $\widetilde{W}_{t_n}$ is a Brownian motion under the probability
measure $\widetilde{P}_n$.
\end{lemma}

\begin{proof}
See appendix (\ref{MiscProofs})
\end{proof}

\begin{theorem}\label{thrm:condExpExponMart}
Assume that 
$$E[\|\beta_{t_n}\|]<\infty$$
Let 
$$\rho\sigma\widehat{\theta}_{t_n}=\widehat{\beta}_{t_n}-\widetilde{c}_{t_n}$$
Then, it follows that the stochastic difference equation
$$\Delta\left(\frac{1}{\zeta_{t_n}}\right)=\frac{1}{\zeta_{t_n}}\widehat{\theta}^*_{t_n}\Delta
\widetilde{W}_{t_n}$$
holds, and its stochastic integral representation is given by
\begin{equation}\label{expMartingaleFUnderRiskNeutral}
\zeta_{t_n}=\exp\{-\sum_{k=0}^{n-1}
\widehat{\theta}^*_{t_k}\Delta\widetilde{W}_{t_k}+\frac{1}{2}\sum_{k=0}^{n-1}\widehat{\theta}^*_{t_k}\rho\widehat{\theta}^*_{t_k} \Delta t  \} 
\end{equation}
\end{theorem}

\begin{proof}
See appendix (\ref{MiscProofs})
\end{proof}

\begin{definition}
The financial market $\mathcal{M}$ is viable, if the following
conditions are satisfied

\begin{equation}\label{Novikov}
P[\omega:\sum_{k=0}^{n-1} \theta^*_{t_k}\rho\theta_{t_k} \Delta s<\infty]=1
\end{equation}
$$E[Z_{t_n}]=1$$
\end{definition}
\begin{remark}
The intuition behind the above argument is, that there is a  price of
risk that the market will quote you, that reflects the market risk
net of transaction costs. In this setting of futures investments, and contrary to the literature, it is economically intuitive to formulate the market price
of risk without the usual risk-free rate term that translates the market
risk to excess market risk. As suggested in remark (\ref{remark-futTrad}), in futures trading, the investor
earns proceeds from the capital twice, through investing both on
risky and risk-less assets concurrently. On the other hand, a long-short strategy
in stocks is cash-intensive, and the same capital has to be either allocated
in the risk-free account or on some risky strategy, which makes
the trade-off between risky and risk-free assets relevant.
\end{remark}

Define the \emph{adjusted discount process} by

\begin{equation}\label{adjDiscountProc}
\gamma_{t_n}=\exp\left\{-\sum_{k=0}^{n-1}(1-m)r \Delta t\right\}
\end{equation}

and the \emph{state price density} by
\begin{equation}\label{statePriceDensity}
H_{t_n}=\gamma_{t_n}Z_{t_n}
\end{equation}
and recall that m (see remark (\ref{remark-futTrad})) is a non-negative constant with compact support in $[0,1]$ that  represents the margin as proportion
of the wealth $X^{\pi,x}_{t_n}$. We also introduce the following notation
$$h_{t_n}=\gamma_{t_n}\zeta_{t_n}$$

\begin{definition}\label{admissibility}
The initial endowment $x>0$ is independent of the
$\sigma$-algebra generated by $F(\cdot)$, and satisfies
$$E[\|x\|^2] <\infty$$
A portfolio process $\pi(\cdot)$ is called admissible if it satisfies
$$x>0 \ \  and \ \  X^{\pi,x}_{t_n}\geq 0 \ \ \forall t \in \mathcal{S}$$
 The set of all
portfolio processes, where the above conditions hold, will be denoted by $\mathcal{A}(x)$
\end{definition}

Assume there is an agent, who can decide on a non-anticipative basis, to
invest her capital at time $t_n$, to the $d$ available futures by selecting
an appropriate vector of positions $P^{\pi}_{t_n}$. We assume that the positions
are sufficiently small, in the sense that they do not "move" the market (see
assumption \ref{maxPos}). Then, the wealth equation of an investor $X^{\pi,x}_{t_n}$
with portfolio process $\widehat{\pi}^c_{t_n}$ and initial endowment $x$, is derived as
follows:
$$\Delta X^{\pi,x}_{t_n}=(1-m)r X^{\pi,x}_{t_n} \Delta t
+\Delta F^*_{t_n}\diag(f)P^{\pi,c}_{t_n}-\frac{1}{2}c^*\diag(f)|\Delta P^{\pi,c}_{t_{n-1}}|$$
$$\Delta X^{\pi,x}_{t_n}=(1-m)r X^{\pi,x}_{t_n} \Delta t
+X^{\pi,x}_{t_n}\Delta F^*_{t_n}\diag(f)\diag(C^{-1}_{t_n})\pi^c_{t_n}-\frac{1}{2}c^*\diag(f) |\Delta P^{\pi,c}_{t_{n-1}}|$$
$$\Delta X^{\pi,x}_{t_n}=(1-m)r X^{\pi,x}_{t_n} \Delta t
+X^{\pi,x}_{t_n}\Delta F^*_{t_n}\diag(F^{-1}_{t_n})\pi^c_{t_n}-\frac{1}{2}c^*\diag(f) |\Delta P^{\pi,c}_{t_{n-1}}|$$
$$\Delta X^{\pi,x}_{t_n}=(1-m)r X^{\pi,x}_{t_n} \Delta t + X^{\pi,x}_{t_n}(\pi^c_{t_n})^*(\beta_{t_n}-\frac{1}{2}
c\left((P^{\pi,c}_{t_n})^*\diag(C_{t_n})\Delta t\right)^{-1}\diag(f)|\Delta P^{\pi,c}_{t_{n-1}}|)\Delta t$$
$$ + X^{\pi,x}_{t_n}(\pi^c_{t_n})^*\sigma \Delta W_{t_n}$$
$$\Delta X^{\pi,x}_{t_n}=(1-m)r X^{\pi,x}_{t_n} \Delta t + X^{\pi,x}_{t_n}(\pi^c_{t_n})^*(\beta_{t_n}-\widetilde{c}_{t_n})\Delta t$$
\begin{equation}\label{wealthProcess}
 + X^{\pi,x}_{t_n}(\pi^c_{t_n})^*\sigma \Delta W_{t_n}
\end{equation}
where 
\begin{equation}\label{eq:marketPosition}
P^{\pi,c}_{t_n}=X^{\pi,x}_{t_n}\diag(C^{-1}_{t_n})\widehat{\pi}^c_{t_n}
\end{equation}
is the futures contacts position.
The "one-way" spread relative to the price is a 
 $\widetilde{c}:\mathcal{S} \times \Omega \rightarrow
\mathbb{R}^d$ process
and is given by 
\begin{equation}\label{eq:costFn}
\widetilde{c}^{(i)}_{t_n}=\frac{c^{(i)}f^{(i)}|\Delta P^{\pi,c(i)}_{t_{n-1}}|}
{2P^{\pi,c(i)}_{t_n}C^{(i)}_{t_n}\Delta t}
\end{equation}
for the ith asset at time t. We denote the vector formulation of $\widetilde{c}$ by $\widetilde{c}_{t_n}=\{\widetilde{c}^1_{t_n},\ldots,\widetilde{c}^d_{t_n}\}$,
and we refer to $|\cdot|$ as the component-wise absolute value. The superscript
in $\pi^c$ denotes the dependence of the portfolio process to non-zero transaction
cost.
\begin{remark}\label{rmk:gearing}
Let $k$ be the gearing/leverage weights that take values in $\mathbb{R}^d_+$. In managed futures funds, the gearing vector is a proprietary vector of weights
that plays the important role of
establishing the leverage of the fund. A typical futures fund pledges to
investors to fluctuate within a certain band of volatility of monetary returns.\footnote{We define monetary returns in the interval $[t_{n-h},t_n],\ \ h \in \mathbb{N}$ by $\frac{1}{X^{\pi,x}(t)}
\sum_{k=n-h}^{n}(P^{\pi,c}_{t_k})^*\diag(f)\Delta F_{t_k}$}
Generally, futures funds tend to market their strategies between 10\%- 25\% of annualised volatility.
\end{remark}

\begin{discussion}

The first term in eq. (\ref{wealthProcess}) refers to the capital earned
by the wealth invested in the interest free account. Note, that we adjust
the wealth by $1-m$ factor, as we assume that we do not earn interest on
the margin.
The second term refers to capital invested in the $d$-dimensional futures
contracts, where $\pi$ refers to the portfolio process.
Finally, the last term is the diffusion term of the wealth equation. Note
that the position $P^{\pi,c}$ can be arbitrarily multiplied by the
leverage term $\kappa$ in order to achieve the required volatility
of returns $\alpha$
$$\var\left( \frac{X^{\pi,x}_{t_n}-X^{\pi,x}_{t_n-\Delta t}}{X^{\pi,x}_{t_n-\Delta t}}\right)=\alpha, \\\ \alpha \in \mathcal{R}_+$$
where $\Delta t$ typically represents a daily time interval.
\end{discussion}

From eq. (\ref{brownianUnderRiskNeutral}) and
(\ref{marketPriceOfRisk}, \ref{wealthProcess}, \ref{brownianUnderRiskNeutral}),
we deduce that the wealth equation permits the
following stochastic representation
$$\Delta X^{\pi,x}_{t_n}=(1-m)r X^{\pi,x}_{t_n} \Delta t + X^{\pi,x}_{t_n}(\pi^c_{t_n})^*(\beta_{t_n}-\widetilde{c}_{t_n})\Delta t$$ $$+ X^{\pi,x}_{t_n}(\pi^c_{t_n})^*\sigma (\Delta\widetilde{W}_{t_n}-\rho\theta_{t_n}\Delta t)$$
\begin{equation}\label{WealthUnderRiskNeutral}
=(1-m)r X^{\pi,x}_{t_n} \Delta t + X^{\pi,x}_{t_n}(\pi^c_{t_n})^*\sigma \Delta\widetilde{W}_{t_n}
\end{equation}

\begin{lemma}
The margin adjusted, state price density portfolio process $$\gamma_{t_n}Z_{t_n}X^{\pi,x}_{t_n}$$
is a local martingale and a supermartingale.

\end{lemma}

\begin{proof}
Note that $\gamma_{t_n}$ is a decreasing
function of $t$. From the monotonicity of $\gamma_{t_n}$ and the Novikov condition (\ref{Novikov}), it follows that $\gamma_{t_n}Z_{t_n}X^{\pi,x}_{t_n}$ is bounded.

Then, by direct application of Taylor theorem and discarding the higher order
terms of $\Delta t$, we obtain
$$\Delta\gamma_{t_n}X^{\pi,x}_{t_n}=-\gamma_{t_n}(1-m)r X^{\pi,x}_{t_n}\Delta t + \gamma_{t_n}(1-m)r
X^{\pi,x}_{t_n}\Delta t$$
$$+ \gamma_{t_n}X^{\pi,x}_{t_n}(\pi^c_{t_n})^*\sigma \Delta\widetilde{W}_{t_n}$$
\begin{equation}\label{discountedWealth}
\gamma_{t_n}X^{\pi,x}_{t_n}=x+\sum_{k=0}^{n-1} \gamma_{t_k}X^{\pi,x}_{t_k}(\pi^c_{t_k})^* \Delta\widetilde{W}_{t_k}
\end{equation}

Now, recall that $H_{t_n}$ is the state price density given by (\ref{statePriceDensity}), and apply Taylor expansion in the
$\Delta H_{t_n}X^{\pi,x}_{t_n}$.
The boundeness of $\Delta H_{t_n}X^{\pi,x}_{t_n}$ follows using similar arguments as
above.
\begin{equation}
\Delta H_{t_n}X^{\pi,x}_{t_n}=(\Delta\gamma_{t_n}X^{\pi,x}_{t_n})Z_{t_n}+(\Delta Z_{t_n})\gamma_{t_n}X^{\pi,x}_{t_n}+\Delta[\gamma_{t_n}X^{\pi,x}_{t_n},Z_{t_n}]
\end{equation}

Note that $\Delta Z_{t_n}=-\theta_{t_n}Z_{t_n}\Delta W_{t_n}$ follows from Taylor's theorem application (we already established that in lemma (\ref{GirsanovLemma})).
$$=H_{t_n}X^{\pi,x}_{t_n}(\pi^c_{t_n})^*\sigma \Delta\widetilde{W}_{t_n}-H_{t_n}X^{\pi,x}_{t_n}\theta_{t_n}\Delta
W_{t_n}$$
$$-H_{t_n}X^{\pi,x}_{t_n}(\pi^c_{t_n})^*\sigma \rho\theta_{t_n}\Delta t$$
$$=H_{t_n}X^{\pi,x}_{t_n}\left(\sigma\pi^c_{t_n}-\theta_{t_n}\right)^*\Delta W_{t_n}$$

or equivalently in Ito's sum form
\begin{equation}\label{stateDensityWealth}
H_{t_n}X^{\pi,x}_{t_n}=x+\sum_{k=0}^{n-1} H_{t_k}X^{\pi,x}_{t_k}\left(
\sigma\pi^c_{t_k}-\theta_{t_k} \right)^*\Delta W_{t_k}
\end{equation}

Eq. (\ref{stateDensityWealth}) is a stochastic integral, and therefore a
local martingale. Subject to the Novikov's condition, and by Fatou's lemma, we deduce that any non-negative local
martingale is a supermartingle.

It follows from above that the budget constraint equation is given by
\begin{equation}\label{budgetConstraint}
E[H_{t_N}X^{\pi,x}_{t_N}]\leq x
\end{equation}
\end{proof}

To ensure text continuity, we refer the auxiliary results
about utility functions stated in the section \ref{sec-UtilFns} of the appendix, and we introduce a special class of normal concave
utilities of stochastic processes, following similar arguments to that of
normal convex integrands of Rockafellar \cite{R68} (see section \ref{subsec:normConcUtil}
in the appendix).



\section{The Optimization Problem}\label{sec:OptimProblem}
\subsection{The Primal  Problem}

Consider an investor who wants to maximize the expected value of his terminal
wealth $E[U(X^{\pi,x}_{t_N})]$ over the class of admissible trading stategies
$\mathcal{A}$ (see def.
(\ref{admissibility})), where  $U:(0,\infty)\rightarrow \mathbb{R}$ is a strictly increasing, strictly concave function, continuously differentiable utility function that satisfies
\begin{equation}\label{inftyCond}
E[U(X^{\pi,x}_{t_N})]<\infty
\end{equation}

We name the objective function under the primal problem by $\mathbf{P}$,
and we denote the value function by
\begin{equation}\label{valueFn}
V(x)=\sup_{\pi \in \mathcal{A}} E[U(X^{\pi,x}_{t_N})]
\end{equation}
subject to
$V(x)<\infty,\ \ \forall x \in (0,\infty)$. The objective function (\ref{valueFn})
is deduced from (\ref{dualFn1}), when --in the notation of section (\ref{sec:auxResults})--,
we evaluate $\widetilde{U}(y)$ at $y=0$. The optimal value of  $\mathbf{P}$
is therefore
$$\inf \mathbf{P} =\inf_{\widetilde{\pi} \in \mathcal{\widetilde{A}}}V(x)$$

The $\mathbf{P}$ can be less
rigorously interpreted to the following
dynamic optimization problem

$$\sup_{\pi \in \mathcal{A}} E[U(X^{\pi,x}_{t_N})]$$

$$s.t. \ is \ financed \ by \ \pi \in  \mathcal{A}(x) \  \& \  with\  the \ initial\  endowment\  being \ x$$

Note that the above optimization problem has infinitely many budget constraints.

A necessary and sufficient (for justification of the assumption, see KLSX\cite{KLSX91} ) condition for the above assumption is

$$U(x)\leq k_1 +k_2 x^\delta, \ \ k_1>0, k_2>0, \delta \in (0,1)$$

A portfolio process $\pi$ that achieves the supremum of the expected value
of the utility function evaluated at the  terminal wealth  value at time
$T$, is called optimal.

\begin{lemma}\label{lemma-ConvValFn}
The value function $V$ (\ref{valueFn}) is a continuous, non-decreasing concave
function.
\end{lemma}

\begin{proof}

The non-decreasing property of $V$ is easily observed by the fact that the
set $\mathcal{A}(x)$ increases as $x$ increases (see def. (\ref{admissibility})).
Now, let $0 \leq p=1-q \leq 1$ and $x_1,x_2>0$. From the concavity of the utility
function $U$, we have

$$p V(x_1) +q V(x_2)$$
$$\leq V(p x_1 + q x_2)$$

i.e. the graph lies above any of its chords. Continuity is implied by concavity.

\end{proof}

\subsection{The Dual Problem}

The dual portfolio process is an
$\mathbb{R}^d$ value process adapted with respect to $\mathcal{F}^R_{t_n}$
that satisfies
$$E[\sum_{k=0}^{n-1} || \widetilde{\pi}_{t_k}||^2 \Delta t] < \infty \ \ a.s. $$

for $\widetilde{\pi}:\mathcal{S} \times \Omega \rightarrow \mathbb{R}^d$.

We name the objective function under the dual problem by $\mathbf{D}$,
and we denote the value function by
$$\widetilde{V}(y)=\inf_{\widetilde{\pi} \in \mathcal{\widetilde{A}}}
 E[U(Y^{\widetilde{\pi},x}_{t_N})]$$

and we assume that
$V(y)<\infty,\ \ \forall y \in (0,\infty)$, where $\mathcal{\widetilde{A}}$
is the admissible class under the dual problem.
The optimal value of  $\mathbf{D}$ is
$$\sup \mathbf{D} =\sup_{\pi \in \mathcal{A}}\widetilde{V}(y)$$

Existence and uniqueness of the dual value function follows from the strict convexity of  $\widetilde{U}(\cdot)$
(see XS\cite{XS92}).

\begin{lemma}
The value function $\widetilde{V}$ (\ref{valueFn}) is a continuous, non-increasing convex function.
\end{lemma}

\begin{proof}

The proof follows from similar arguments as lemma (\ref{lemma-ConvValFn}).

\end{proof}

Moreover, the following assumption (see \cite{KS99}) ensures existence of the optimal solution.

\section{Optimal Portfolio Process}\label{sec:OptimPortf}
In the following section we present our main result. We show that by solving
the static optimization problem (see section \ref{subsec:static} for a similar proof
to \cite{CH88} that the solution to the static problem coincides to the solution
of the dynamic problem) we obtain an optimal portfolio process characterization. 

\subsection{Portfolio Process Characterization}

In the following section, we prove that the dual portfolio process $\widetilde{\pi}$
provides a bound for the value function V. We follow similar arguments
to KLSX\cite{KLSX91}, CK\cite{CK92}
to show the existence of the optimal $\pi(\cdot)$ under the primal problem.

We introduce the following mapping $\mathcal{X}:(0,\infty) \rightarrow (0,\infty)$,
and we define

\begin{equation}\label{bigX}
\mathcal{X}(y)=E\left[H_{t_N}I(yH_{t_N})  \right] , \ \ \   0<y<\infty
\end{equation}

Note, that the monotonicity(strictly decreasing) and continuity properties
of $\mathcal{X}(y)$
follow directly from properties of $I(\cdot)$. Moreover, define $\mathcal{Y}(\cdot)$
as the inverse of $\mathcal{X}(\cdot)$.

\begin{theorem}\label{optimPorfExist}
Let
\begin{equation}\label{optimalTermWealth}
\xi=I(\mathcal{Y}(x)H_{t_N})
\end{equation}
and assume that satisfies the following condition
$E[U(\xi))]< \infty$.
The value function $V$ attains its supremum if and only if
$$E[H_{t_N}\xi]=x$$
Then, $\forall \pi \in \mathcal{A}(x)$, we have
$$V(x) \leq E[U(\xi)]$$
\end{theorem}
\begin{proof}
See appendix (\ref{MiscProofs})
\end{proof}

From theorem (\ref{optimPorfExist}) follows that there is a portfolio process
$\widehat{\pi}^c$ that attains the supremum of the  value function, and we
call it optimal when we have the following equality $X^{\widehat{\pi},x}_{t_N}=\xi_{t_N}=\xi$,
for $\xi$ defined in eq. (\ref{optimalTermWealth}).

The following lemma is in the spirit of \cite{CK92,KLS87,KLSX91,XS92}.
\begin{lemma}
The optimal portfolio process $\widehat{\pi}$ is
defined by

\begin{equation}\label{optimalPortf}
\widehat{\pi}^c_{t_n}=\frac{1}{h_{t_n}\xi} (\sigma)^{-1}
(g_{t_n}+M_{t_n}\widehat{\theta}_{t_n})
\end{equation}
\end{lemma}

\begin{proof}

Define the positive martingale
$$M(t)=E[h_{t_N}X^{\widehat{\pi},x}_{t_N}|\mathcal{F}_{t_n}]$$

Then, there exists a unique stochastic process $g(s,\omega)$, such that
$M_{t_n}$ admits the following
stochastic integral representation (see thrm 4.3.4 of Oksendal\cite{O06})
$$M_{t_n}= x+ \sum_{k=0}^{n-1} g^*_{t_k}\Delta W_{t_k} , \ \ a.s  \forall t\geq0$$

and the mapping $g:\mathcal{S} \times \Omega \rightarrow \mathbb{R}^d$, satisfies $\sum_{k=0}^{n-1} ||g_{t_k}||^2\Delta t<\infty$.

Firstly, note that from Taylor theorem and eq. (\ref{brownianUnderRiskNeutral}),
we have
$$\Delta \frac{1}{\zeta_{t_n}}=\frac{1}{\zeta_{t_n}}\widehat{\theta}^*_{t_n}\Delta \widetilde{W}_{t_n}$$
where $\widehat{\theta}_{t_n}=(\rho\sigma)^{-1}\widehat{\beta}_{t_n}$.
By Taylor expansion, we obtain
$$\Delta \frac{1}{\zeta_{t_n}}M_{t_n}=\left(\Delta \frac{1}{\zeta_{t_n}}\right)M_{t_n}+(\Delta M_{t_n})\frac{1}{\zeta_{t_n}}+\Delta [\frac{1}{\zeta_{t_n}},M_{t_n}]$$
$$\Delta \frac{1}{\zeta_{t_n}}M_{t_n}=\frac{1}{\zeta_{t_n}}M_{t_n}\widehat{\theta}^*_{t_n}\Delta \widetilde{W}_{t_n}+
\frac{1}{\zeta_{t_n}} \left( g^*_{t_n}\Delta \widetilde{W}_{t_n}-g^*_{t_n}\rho\widehat{\theta}_{t_n}\Delta t \right)$$
$$+\frac{1}{\zeta_{t_n}}g^*_{t_n}\rho\widehat{\theta}_{t_n}\Delta t$$
\begin{equation}
=\frac{1}{\zeta_{t_n}} \left( g_{t_n}+ M_{t_n} \widehat{\theta}_{t_n}  \right)^* \Delta \widetilde{W}_{t_n}
\end{equation}

Consequently, the discounted optimal wealth is defined as
\begin{equation}\label{discountedWealth2}
\gamma_{t_N}\xi=x+ \sum_{k=0}^{n-1}\frac{1}{\zeta_{t_k}} \left( g_{t_k}+ M_{t_k} \widehat{\theta}_{t_k}  \right)^* \Delta \widetilde{W}_{t_k}
\end{equation}
Substitute the optimal portfolio process (\ref{optimalPortf})
in (\ref{discountedWealth})  to obtain the portfolio process given
by (\ref{discountedWealth2}). That completes the proof.

\end{proof}

\section{Examples}\label{sec:Examples}

Consider the case, where the utility of the value function (\ref{valueFn})
is logarithmic $U(x)=log(x)$.
Then, according to theorem (\ref{optimPorfExist}), the explicit formulation of the terminal wealth that maximizes the value function is

\begin{equation}\label{optimWealthLogFn}
\xi=x \exp \left\{ \sum_{k=0}^{n-1} (1-m)r+\frac{1}{2}\widehat{\theta}^*_{t_k}\rho\widehat{\theta}_{t_k} \Delta t + \sum_{k=0}^{n-1} \widehat{\theta}^*_{t_k}\Delta W_{t_k}  \right\}
\end{equation}
where $\mathcal{X}(y)=\frac{1}{y}$ and  $\mathcal{Y}(x)=\frac{1}{x}$.

Now, rearranging (\ref{stateDensityWealth}),
\begin{equation}\label{wealthEqn}
X^{\pi,x}_{t_n}=\frac{x}{H_{t_n}}+\frac{1}{H_{t_n}}\sum_{k=0}^{n-1} H_{t_k}X^{\pi,s}_{t_k}\left(
\sigma\pi^c_{t_k}-\theta_{t_k} \right)^*\Delta W_{t_k}
\end{equation}
and substituting the optimal portfolio process (\ref{optPortf-LogUtil})
$$\widehat{\pi}^c_{t_n}=(\sigma)^{-1}\widehat{\theta}_{t_n}$$
\begin{equation}\label{optPortf-LogUtil}
\widehat{\pi}^c_{t_n}=(\sigma\rho\sigma)^{-1}(\widehat{\beta}_{t_n}-\widetilde{c}_{t_n})
\end{equation}
\begin{equation}\label{optPortfPos-LogUtil}
P^{\widehat{\pi},c}_{t_n}=X^{\widehat{\pi},x}_{t_n}
\diag(C^{-1}_{t_n})\widehat{\pi}^c_{t_n}
\end{equation}
into (\ref{wealthEqn}), we conclude that
$$X^{\widehat{\pi},x}_{t_n}=\frac{x}{h_{t_n}}=\xi$$

Therefore, the optimal portfolio process is given by eq. (\ref{optPortf-LogUtil})
and the value function is

$$V(x)=E[\log X^{\widehat{\pi},x}_{t_N}]=\log x +\sum_{k=0}^{n-1} (1-m)r+\widehat{\theta}^*_{t_k}\rho\widehat{\theta}_{t_k}  \Delta t $$
Equivalently, when the bid-ask spread is zero, we have
\begin{equation}\label{optPortf-LogUtil_ZeroTrCost}
\widehat{\pi}_{t_n}=(\sigma\rho\sigma)^{-1}\widehat{\beta}_{t_n}
\end{equation}
\begin{equation}\label{optPortfPos-LogUtil_ZeroTrCost}
P^{\widehat{\pi}}_{t_n}=X^{\widehat{\pi},x}_{t_n}
\diag(C^{-1}_{t_n})\widehat{\pi}_{t_n}
\end{equation}
which is simply the formulation of the optimal portfolio process/position without
transaction costs.
\begin{remark}
Substituting the optimal portfolio process (\ref{optPortf-LogUtil}) into the position process (\ref{eq:marketPosition}), and introducing a scalar
multiplier, the gearing process (see remark (\ref{rmk:gearing})), we have
$$\widetilde{P}^{\widehat{\pi},c}_{t_n}=X^{\widehat{\pi},x}_{t_n}\diag(k)\diag(C^{-1}_{t_n})\widehat{\pi}^c_{t_n}$$
Moreover, in order to retain the positivity of the expected payoff of the investor,
we need to satisfy the following condition
$$|\widehat{\beta}|-|\widetilde{c}|\geq0$$
or equivalently to introduce
the following transformation
\begin{equation}\label{eq:payoff}
\Upsilon_{t_n}=\max(\widehat{\beta}_{t_n}-\widehat{c}_{t_n},0)
+\min(\widehat{\beta}_{t_n}-\widehat{c}_{t_n},0)
\end{equation}
and hence, we obtain the following position and portfolio process
\begin{equation}\label{eq:optimalMarketPosition}
\widetilde{P}^{\widehat{\pi},\widehat{c}}_{t_n}=X^{\widehat{\pi},x}_{t_n}\diag(k)\diag(C^{-1}_{t_n})\widehat{\pi}^{\widehat{c}}_{t_n}
\end{equation}
\begin{equation}\label{eq:optimalMarketProcess}
\widehat{\pi}^{\widehat{c}}_{t_n}=(\sigma\rho\sigma)^{-1}\Upsilon_{t_n}
\end{equation}
by using the following approximation for the cost term
\begin{equation}\label{eq:costTermApprox}
\widehat{c}^{(i)}_{t_n}=\frac{c^{(i)}f^{(i)}|\widetilde{P}^{\widehat{\pi}(i)}_{t_n}- \widetilde{P}^{\widehat{\pi},\widehat{c}(i)}_{t_{n-1}}|}
{2\widetilde{P}^{\widehat{\pi}(i)}_{t_n}C^{(i)}_{t_n}\Delta t}
\end{equation}
The final position functional (eq. \ref{eq:optimalMarketPosition}) makes
use of the approximating function $\widehat{c}$ of $\widetilde{c}$, where
$ \widetilde{P}^{\widehat{\pi},c}_t$ is approximated by  $\widetilde{P}^{\widehat{\pi}}_t$,
the optimal position process under logarithmic utility without transaction
costs.

The position process $\widetilde{P}^{\widehat{\pi},\widehat{c}}_{t_n}$ is interpreted as the number of positions for each asset $i$ subject to the gearing vector $k$, in order the portfolio to achieve the required volatility of monetary returns (see section (\ref{subSec:InvProcess}) for further explanation). The vector $k$ could also be interpreted as a vector of proprietary portfolio weights.
\end{remark}

\begin{discussion}
The optimal portfolio/position process in (\ref{eq:optimalMarketPosition}) is a classic
mean-variance criterion, with the difference that our $\widetilde{P}^{\widehat{\pi},\widehat{c}}_{t_n}$ mapping
refers to the d-dimensional position we take in the futures markets. Moreover,
contrary to the literature that reports $\pi$ as a proportion, in our context
it represents an arbitrary weight, and moreover we note the introduction of few
additional terms, namely the bid-ask spread, the contract size, the gearing,
the correlation matrix, and the futures unit value(compare to \cite{KLSX91} who derives the usual mean-variance criterion). 

The incorporation of transaction costs, has a remarkable effect in the position.
We observe that the $\widehat{\beta}_{t_n}$ term in (\ref{optPortf-LogUtil}) is adjusted by another process, namely $\widetilde{c}_{t_n}$, that corresponds
to the cost of entering that position. The latter process is simply the slippage
(see def. (\ref{slippage})) incurred in the market proportional to our expected
movement. We explicitly assume a linear relation between  slippage and magnitude
of market order, but that is again reasonable subject to the assumption (\ref{maxPos})(See
\cite{KW02} for non-linear formulations).

We need, additionally,
to observe that at infitesimal time changes, our expectation will hardly
change but the cost of purchasing/selling a contract is invariant of time
changes (note that we only divide the cost term by $\Delta t$).
\end{discussion}

\begin{remark}Assume that the net return process $\Upsilon_{t_n}$ in (\ref{eq:payoff}) represents our expectation of a particular trading strategy for asset $i$. Then, using Chebyshev inequality, we have that
$$P(\Upsilon_{t_n}>\epsilon) \leq \frac{E[\Upsilon_{t_n}]}{\epsilon}, \ \ \epsilon>0$$
for significantly small $\epsilon$.
But as $\Pi \rightarrow 0$, we assert that $\Upsilon_{t_n}=0$ i.e.
$$P(\Upsilon_{t_n}>\epsilon)=0$$
unless the increment $\Delta t$ is sufficiently large such that $|\beta_{t_n}|\geq \widetilde{c}_{t_n}$.
\end{remark}

\section{Conclusion}\label{sec:Conclusion}

We studied the problem of optimal investment under transaction
cost and offered a number of results pertaining a relative arbitrage trading strategy. Our study focuses in futures markets, contrary to the literature where the bulk of the studies
in optimal investment are in equities markets and few recent studies under
transaction costs in currency markets (see \cite{CPT99,KL02}).

The application of optimal investment strategies in futures markets necessitates
the introduction of concepts such as position, gearing/leverage, slippage and margin, and the modelling of those in the wealth process.  Moreover,
we  offer a brief description of the orderbook, and how it associates to the slippage used in the model under some boundness criteria.

We show explicitly, through the construction of the partial information dynamics, how the filtration generated by the futures prices is equal to the filtration generated
by the return process. We introduce the concept of relative arbitrage in systematic trading, and we offer some intuitive remarks how is related to the innovation process and why such strategy may be profitable.

We proceed by constructing the wealth process by including transaction
costs, and deploying the futures contracts position as control. These formulae account for the
actual transaction cost, rather than deploying proportional constant
transaction costs to the wealth process. Then, we use an arbitrary gearing factor to achieve the
required rate of volatility the fund strategy entails.

We introduce the normal concave utilities, in line with Rockafellar \cite{R68}, whose paper provides a detailed exposition of integrals which are convex functionals. Under this very generic class of utilities, we formulate the
primal and dual problem, and provide conditions under which there is no duality
gap. We continue using arguments adapted from \cite{CK92,KLS87,KLSX91} to show
that the value function attains its supremum under the static problem, which
in turn proves to be equivalent to the dynamic. We conclude the section by
providing an example using logarithmic utility, and an explicit representation
of the optimal portfolio process.

The above theory could be readily extended to the constrained case, in the
spirit of the work of  \cite{C97,CK92}, as the position could be constrained due to liquidity issues. Moreover, the above analysis of the trading strategy could encompass results in stochastic filtering theory to find estimators for the drift of the wealth process, and to study the properties of the estimator using statistical inference tools for stochastic processes.
In addition, we could model the orderbook using stochastic processes and
relax the assumption \ref{maxPos} that relates to remark \ref{remark:orderbook}.

\section*{Acknowledgements}

I would like to thank Aleksandar Mijatovic for helpful comments and suggestions.

\newpage
\appendix{APPENDIX}
\vspace{1cm}

\section{Auxiliary Results}\label{sec:auxResults}

\subsection{Normal Concave Utilities}\label{subsec:normConcUtil}

In this section, we provide some results on concave utilities functionals evaluated at some random process realization for some elementary event $\omega$, to serve as auxiliary
results for establishing existence and uniqueness of solutions.

\begin{definition}\label{normalIntCond}
A concave utility $U(X_t)$, defined on $[0,\infty)$ and maps to $\mathbb{R}_+$,is normal if it is proper (i.e. epigraph is non-empty
and contains no vertical lines) and is continuous, and if there is a family
$E$ of random variables $X$ from $[0,T] \rightarrow \mathbb{R}$
that satisfy:

\begin{itemize}
\item $\forall X_t \in E$, $(X_t,\mathcal{F}_t)$  is progressively measurable
process i.e. for every
time $t$, the map $[0,T] \times \Omega$ is $\mathcal{B}([0,T]) \bigotimes
\mathcal{F}$ measurable and adapted to $\mathcal{F}_t$ .
\item $\forall t \in [0,T]$, $E\bigcap \dom U$ is dense in $\dom U$
\item $\forall t \in [0,T]$, $U$ is essentially smooth, and satisfies the
following conditions for $C=int(\dom U))$
\begin{enumerate}
\item $C$ is not empty
\item $U$ is differentiable everywhere in $C$
\end{enumerate}
\end{itemize}

\end{definition}

\begin{lemma}
Let $U$ be a continuous, proper concave functional in $\mathbb{R}_+$. Then,
$U$ is a normal concave utility.
\end{lemma}

\begin{proof}
Since the effective domain of $U$ is a nonempty concave set, we are able
to define $D$ as a dense subset of $\dom U$. Moreover, let $E$ be the family of random variables $X$ from $[0,T] \rightarrow \mathbb{R}$. Then,
the conditions stated in def. (\ref{normalIntCond}) are satisfied.
\end{proof}

\begin{lemma}
Suppose $U$ is continuous in $X$ for each t, and for each $X$ is measurable
in $t$, and has interior points in its effective domain. Then, $U$ is a normal
concave utility.
\end{lemma}

\begin{proof}
Let the the family $E$ of random variables take values in the dense subset $D$. Then,
$U$ is continuous in $X$ , because $D$ has dense intersection with $\dom U$,
and the measurability condition holds by hypothesis.
\end{proof}

\begin{lemma}
Let $\widetilde{U}$ the conjugate of the normal concave utility $U$.Then,
for every progressively measurable process $X_t$ from $[0,T] \rightarrow \mathbb{R}$, the
function $\widetilde{U}(X_t)$  is measurable.
\end{lemma}

\begin{proof}
We have by definition that

$$-\widetilde{U}(y)=\inf \{ <X_t,y> - U(X_t) |y \in \mathbb{R}\}$$
where the notation "$<>$" here denotes inner product.
In the above formulation, we only need to show that any value of
$U(X_t)=<X_t,y>$ could be approximated by values in $E\bigcap \dom
U$. By hypothesis, we have that $E\bigcap \dom U$ is dense and moreover given
that the $\dom U$ is the closure of the relative interior, the intersection
of $E$ with the relative interior must be dense. According to \cite{R96},
$U$ is continuous in the relative interior of the effective domain $U$, and
the values at boundary points can be obtained as limits. Hence, the values
of $U$  are limits of those for $ E\bigcap \dom U$.

It follows that the pointwise infimum is a collection of measurable functions,
and therefore measurable.

\end{proof}

\subsection{Utility Functions}\label{sec-UtilFns}

A utility function $U$ is a strictly increasing, \textbf{normal} concave mapping that satisfies the Inada conditions
\begin{equation}\label{eqn:Inada}
U'(0+)=\lim_{x\rightarrow0}U'(x)=\infty,
\ \ \
U'(\infty)=\lim_{x\rightarrow \infty}U'(x)=0
\end{equation}
and the following growth condition
\begin{equation}\label{eqn:UtilGrowthCond}
0 \leq U(x) \leq \alpha(1+x^\nu)
\end{equation}
for $\alpha>0$ and $\nu \in (0,1)$.

Denote by $I:(0,\infty)\rightarrow(0,\infty)$ the strictly decreasing, continuous inverse of $U'$. This function satisfies
$$I(0+)=\infty,\ \  I(\infty)=0$$
and
$$U'(I(y))=y \ \forall y>0, \ \ \ U'(I(x))=x \ \forall x>0 $$

Let $\mathcal{T}$ and $\widetilde{\mathcal{T}}$ be two Hausdorff topological vector spaces separated by a closed affine hyperplane $\mathcal{H}$,  which
can be considered as bilinear form over $\mathcal{T} \times \widetilde{\mathcal{T}}$. Let  $C \subset \mathcal{T}$  be a convex, closed and non-empty set and denote by $D$ the dual  of $C$ (see Hahn-Banach
Theorem in Ekeland, Temam \cite{ET79} for existence of non-zero continuous
linear forms in $\mathcal{T}$). This is to say that
any continuous function of $x$ can be expressed as a linear function of $<x,y>$
and conversely.

\begin{lemma}\label{lemma-legendre}
Let $U$ be a \textbf{normal} concave utility. Then the Legendre conjugate
is defined by
\begin{equation}\label{legendre}
\widetilde{U}(y)=U(I(y))-I(y)y
\end{equation}

Hence, the Legendre conjugate $(D,\widetilde{U})$ of $(C,U)$ is well defined, where $D$ is the image of $C$ under the gradient mapping $\bigtriangledown U$.
\end{lemma}

\begin{proof}
The subdifferential mapping $\vartheta U$ reduces to the gradient mapping
$\bigtriangledown U$, if $U$ is smooth. We know that $U$ is smooth by hypothesis that is normal concave utility. Therefore, for a given $x^*=\bigtriangledown
U$, the value $U(x)-xy$ attains its supremum at $\widetilde{U}(x^*)$. Therefore,
the formula (\ref{legendre}) is equal to $\widetilde{U}(x^*)$.
\end{proof}

Let $U$ be a normal concave utility in $\mathcal{T}$ and $\alpha \in \mathbb{R}$. Then,
the continuous affine function $x \rightarrow \alpha -xy$ dominates $U(x)$
if and only if
$$\forall x \in \mathcal{T}, \ \ \alpha \geq U(x) -xy$$
$$\alpha \geq \widetilde{U}(y)$$

Hence, we proceed
to define the conjugate (or polar) of $U$ by

\begin{equation}\label{dualFn1}
\widetilde{U}(y)=\sup_{x>0}\{ U(x)-xy\}, \ \ y \in \mathbb{R^+}
\end{equation}
which can be seen as the Lagrangian associated with the system of perturbations
$y \in \mathbb{R^+}$.
According to Rockafellar, in order that $\widetilde{U}$ is proper and $\widetilde{\widetilde{U}}=U$,
it is necessary and sufficient that $U$ is lower-semi continuous with respect
to the weak topology induced on $C$ by $D$, but we have already established
continuity through the assumption that $U$ belongs to the class of
normal concave utilities.

From  (\ref{dualFn1}), we obtain
\begin{equation}\label{dualFn2}
U(x)=\inf_{y>0}[\widetilde{U}(y)+xy], \ \ x \in \mathbb{R^+}
\end{equation}

From eqn. (\ref{dualFn1}, \ref{dualFn2}) and lemma (\ref{lemma-legendre}), we obtain the following inequalities

\begin{equation}\label{conjIneq}
U(I(y))\geq U(x)+y[I(y)-x], \forall x>0,y>0
\end{equation}
\begin{equation}
\widetilde{U}(U'(x))\leq \widetilde{U}(y)-x[U'(x)-y], \forall
x>0,y>0
\end{equation}

\newpage

\section{Static Optimization problem}\label{subsec:static}

The set of wealth $X^{\pi,x}(\cdot)$ financed by $\pi$ and having initial
endowment $x$ is given by

$$\mathcal{V}\equiv\{\pi \in \mathcal{A}:  (X^{\pi,x}) \in \mathcal{L}_+^p(P) \}$$
where $\mathcal{L}_+^p(P) \equiv \mathcal{L}_+^p(\Omega,\mathcal{F},P)$
and $0 < p < \infty $.

Now, consider the following static variational problem

$$\sup_{\pi \in \mathcal{A}} E[U(\xi)]$$
subject to
$$E[H_{t_N}\xi]= x$$

In the following proposition, we show that by solving the static problem,
we obtain the solution to the dynamic problem (\ref{valueFn}), and we prove equivalence by contradiction.
The argument is that an agent prefers a $\pi$ solution to the dynamic problem,
whose budget constraint is satisfied. But if the solution is budget feasible, then it
will coincide to the solution  of the static problem.

\begin{proposition}
Consider $\xi$ is a solution to the static variational problem and takes values in $\mathcal{V}$. Then $\xi$ is  the solution to the  dynamic problem. The
converse also holds.
\end{proposition}

\begin{proof}
Assume there exists a solution  $\widehat{\xi} \in \mathcal{V}$ financed by $\pi$ and
initial capital $x$ such that the following inequality holds
$$E[U(\widehat{\xi})]>E[U(\xi)]$$
But from (\ref{budgetConstraint}), we have
$$E[H_{t_N}\widehat{\xi}] \leq x$$
and deduce that $\widehat{\xi}$ is budget feasible, which contradicts the hypothesis that $\xi$ is a solution to the static problem.
The second statement follows from elementary arguments.
\end{proof}

\newpage

\section{Misc. proofs}\label{MiscProofs}
\begin{proof}(lemma \ref{GirsanovLemma})

The proof will follow in two stages. First, we prove that $Z_{t_n}$ is a
martingale/supermartingale and that $E[Z_{t_n}]=1$, and then
we show that $P_n<<\widetilde{P}_n$(absolutely  continuous).
\begin{itemize}

\item By Taylor expansion applied on eq. (\ref{expMartingale}) and discarding
the higher order terms of $\Delta t$, we get

\begin{equation}\label{V-Martingale}
Z_{t_n}=1-\sum_{k=0}^{n-1} Z_{t_k}\theta^*_{t_k}\Delta W_{t_k}
\end{equation}

which is a local martingale.
Let an increasing sequence of Markov times $\tau_n$, such as the sequence
$(Z(t_n\bigwedge\tau_n),\mathcal{F}_{t_n})$ is uniformly integrable, and
$Z(t_n\bigwedge\tau_n)$ satisfies $Z_{t_{n-1}}=E[Z_{t_n}|\mathcal{F}_{t_{n-1}}]$. Then,
from monotone convergence and Fatou's lemma, we obtain that $Z_{t_n}$ is a
 supermartingale.
 The assertion that $E[Z_{t_n}]$=1, follows from eq. (\ref{V-Martingale}), subject
 to the initial condition $Z_{t_0}=1$.
 
By application of Taylor theorem on $\Delta(\widetilde{W}_{t_n}Z_{t_n})$, we have
$$\Delta (\widetilde{W}_{t_n}Z_{t_n})=\widetilde{W}_{t_n}\Delta Z_{t_n}+Z_{t_n} \Delta\widetilde{W}_{t_n}
+\Delta[\widetilde{W}_{t_n},\Delta Z_{t_n}]$$
$$=-\widetilde{W}_{t_n}Z_{t_n}\theta^*_{t_n}\Delta W_{t_n}+Z_{t_n}\Delta\widetilde{W}_{t_n}+
Z_{t_n}\rho\theta_{t_n}\Delta t-Z_{t_n}\rho\theta_{t_n}\Delta t$$
$$=-(\theta^*_{t_n}\widetilde{W}_{t_n}+1)Z_{t_n}\Delta W_{t_n}$$
which is a local martingale.
Therefore,
$$\widetilde{E}[\widetilde{W}_{t_n}|\mathcal{F}^R_{t_{n-1}}]
=\frac{1}{Z_{t_{n-1}}}E[\widetilde{W}_{t_n}Z_{t_n}|\mathcal{F}^R_{t_{n-1}}]
=\widetilde{W}_{t_{n-1}}$$
which proves that $\widetilde{W}_{t_n}$ has the martingale property.

 \item Subject to the assumption $Z(t_0)=1$ i.e. $Z_{t_n}=1$,
 we have that for any measurable set $A \in \mathcal{F}_{t_n}$ ,
 we have from Radon-Nikodym that

 \begin{equation}\label{Radon-Nikodym}
 \mu_W(A)=\widetilde{P}_n(\omega:R \in A)=\int_{\omega:R \in A} Z_{t_n}(\omega) dP_n=\int_{\omega:R \in A} E[Z_{t_n}(\omega)|\mathcal{F}^R_{t_n}] dP_n
 \end{equation}

 Let the random process $\eta_{t_n}(\omega)=E[Z_{t_n}(\omega)|\mathcal{F}^R_{t_n}]$  satisfy the Kolmogorov criterion
 $$E[|\eta_{t_n+\Delta t}-\eta_{t_n}|^\alpha]\leq C|\Delta t|^{1+\varepsilon}$$
 for some given constants $\alpha >0$ and $\varepsilon>0$.

 Then, $\eta_{t_n}(\omega)$ has a modification $\Psi(\eta_{t_n}(\omega))$, that is progressively measurable with respect to $\mathcal{F}^R_{t_n}$, which we denote by
$$\Psi(\eta_{t_n}(\omega))=E[Z_{t_n}(\omega)|\mathcal{F}^R_{t_n}]$$
and we rewrite the last part of eq. (\ref{Radon-Nikodym}) as follows
$$\int_{\omega:R \in A} \Psi(\eta_{t_n}(\omega)) dP_n=\int_{A} \Psi(y) d\mu_R(y)$$
and by (\ref{Radon-Nikodym}), we have 
$$\mu_W(A)=\int_{A} \Psi(y) d\mu_R(y)$$
Hence, we obtain the absolute continuity of $\mu_W$ and $\mu_F$
$$\frac{\mu_W}{\mu_R}=\widetilde{\eta}_{t_n}(\omega)$$
\end{itemize}
\end{proof}

\vspace{1cm}
\begin{proof}(theorem \ref{thrm:condExpExponMart})
We follow similar arguments to \cite{L98}.

From eq. (\ref{expMartingale}) and (\ref{brownianUnderRiskNeutral}) it follows
\begin{equation}\label{expMartingale-riskNeutralMeas}
Z_{t_n}=\exp\{-\sum_{k=0}^{n-1}
\theta^*_{t_k}\Delta \widetilde{W}_{t_k}+\frac{1}{2}\sum_{k=0}^{n-1} \theta^*_{t_k}\rho\theta_{t_k}\Delta
t  \}
\end{equation}
By Taylor expansion theorem, and by discarding the higher order terms
of $\Delta t$, we get
\begin{equation}\label{expMartingaleSDE-riskNeutralMeas}
\Delta\left(\frac{1}{Z_{t_n}}\right)=\zeta_{t_n}\theta^*_{t_n}\Delta \widetilde{W}_{t_n}
\end{equation}
Making use of the assumption that we can replace the quadratic variation
of $\Delta W$ by its expectation, we have
$$[\frac{1}{Z},W]_{t_n}=\sum_{k=0}^{n-1}Z_{t_k}\theta_{t_k}\Delta t$$
From the assumption above and $E[\|\beta_{t_n}-\widetilde{c}_{t_n}\|]<\infty$, it follows
that $\Delta\left(\frac{1}{Z_{t_n}}\right)$ permits the following representation
(see theorem 5.13 in \cite{LS00a})
$$\widetilde{E}[\frac{1}{Z_{t_n}}|F_{t_n}^R]
=\widetilde{E}[\frac{1}{Z_{t_0}}]+\sum_{k=0}^{n-1}\widetilde{E}[Z_{t_k}\theta_{t_k}
|F_{t_k}^R]\Delta\widetilde{W}_{t_k}$$
$$\frac{1}{\zeta_{t_n}}
=1+\sum_{k=0}^{n-1}\zeta_{t_k}\widehat{\theta}_{t_k}
\Delta\widetilde{W}_{t_k}$$
\end{proof}

\vspace{1cm}
\begin{proof}(theorem \ref{optimPorfExist}). The proof follows similar arguments
to \cite{KLS87}, and is restated here only for continuity reasons.

From the budget constraint inequality (\ref{budgetConstraint}) and  (\ref{conjIneq}),
we have
$$U(\xi) \geq U(1) + \mathcal{Y}(x)H_{t_N}(\xi-1)$$
$$\geq -|U(1)| - \mathcal{Y}(x)H_{t_N}$$

The boundness condition follows from the monotonicity of $\gamma_{t_N}$ and the supermartingale property of $Z_{t_N}$.

Moreover, we have
$$U(\xi) \geq U(X^{\widetilde{\pi},x}_{t_N}) + \mathcal{Y}(x)H_{t_N}(\xi-X^{\widetilde{\pi},x}_{t_N})$$
Therefore, from the above inequality, we obtain the required result
$$E[U(\xi)] \geq V(x)$$

The same result can also be obtained from inequality

\begin{equation}\label{conjIneq2}
U(x) \leq \widetilde{U}(y)+xy, \ \ \ \forall x,y >0
\end{equation}
which is derived by rearranging   (\ref{dualFn1}).
Consequently, from  (\ref{budgetConstraint}) and the inequality
(\ref{conjIneq2}), we have
\begin{equation}\label{ineq:OptimInvLagrangian}
U(X^{\widetilde{\pi},x}_{t_N}) \leq \widetilde{U}(\mathcal{Y}(x)H_{t_N})+\mathcal{Y}(x)H_{t_N}X^{\widetilde{\pi},x}_{t_N} \end{equation}
Therefore,
$$V(x) \leq \widetilde{V}(y) +\mathcal{Y}(x)E[H_{t_N}X^{\widetilde{\pi},x}_{t_N}]$$
$$\leq \widetilde{V}(y) +\mathcal{Y}(x)x$$
The above inequality turns to equality, if and only if  (\ref{optimalTermWealth})
holds. The optimality of $\pi^c$ is implied by the existence of optimal solution
for $\widetilde{\pi}^c$.

\end{proof}

\end{document}